\documentclass[12pt,a4paper]{article}
\usepackage{amssymb}

\def\beq{\begin{equation}}
\def\eeq{\end{equation}}

\def\be{\begin{displaymath}}
\def\ee{\end{displaymath}}

\newtheorem{thm}{Result}

\def\Tr{\mathop{\rm Tr}\nolimits}
\def\x {\stackrel {\textstyle \otimes}{,}}
\def\rank{\mathop{\rm rank}\nolimits}
\def\dim{\mathop{\rm dim}\nolimits}

\title{
R-Matrices and Generalized Inverses}

\author{H. W. Braden\\
\normalsize
\em Department of Mathematics and Statistics,\\
\normalsize
\em The University of Edinburgh, \\
\normalsize
\em Edinburgh, UK \\
\normalsize
e-mail: hwb@ed.ac.uk
\\
}

\begin{document}

\renewcommand{\thepage}{}
\begin{titlepage}

\maketitle
\vskip-9.5cm
\hskip10.4cm
MS-97-006
\vskip.2cm
\hskip10.4cm
\sf solv-int/9706001 \rm
\vskip8.8cm

\begin{abstract}
Four results are given that address the existence, ambiguities
and construction of a classical R-matrix given a Lax pair.
They enable the uniform construction of R-matrices
in terms of any generalized inverse of $ad L$. For
generic $L$ a generalized inverse (and indeed the Moore-Penrose
inverse) is explicitly constructed. The R-matrices are in general
momentum dependent and dynamical. The construction 
applies equally to Lax matrices with spectral parameter.
\end{abstract}
\vfill
\end{titlepage}
\renewcommand{\thepage}{\arabic{page}}

\section{Introduction}

The modern approach to completely integrable systems is in terms of Lax
pairs  $L$,  $M$ and R-matrices. Here the consistency of the matrix equation 
$\dot L =\left[ L,M\right] $ expresses the equations of motion of the 
system under consideration. 
The great merit of this approach is that it provides a unified
framework for treating the many disparate completely integrable systems
known.
Given a $2n$-dimensional phase space Liouville's 
theorem \cite{Liouv,Arnold}, which ensures the
existence of action-angle variables, requires that we have $n$ independent
conserved quantities in involution; that is they
mutually Poisson commute.
As a consequence of the Lax equation the traces $\Tr{L\sp k}$ are conserved
and these are natural candidates for the action variables of Liouville's 
theorem.
(In practice the action variables are typically transcendental functions of
these traces.)
It remains however to verify that these traces provide enough independent
quantities in involution. Verifying the number of independent 
quantities is usually straightforward and the remaining step is
then to show they mutually Poisson commute.
The final ingredient of the modern approach, the R-matrix \cite{STS}, 
guarantees their
involution. If $L$ is in a representation $E$ of a Lie algebra 
${\mathfrak{g}}$ (here taken to be semi-simple), the classical 
R-matrix is a $E{\otimes }E$ valued matrix such that
\beq
[R, L\otimes 1]-[{R\sp{T}},1\otimes L] =\{L\x L\}.
\label{rmatrix}
\eeq
(The notation is amplified below.)
Then
$$\{\Tr_E {L\sp k},\Tr_E {L\sp m}\}=
  \Tr_{E\otimes E}\{ {L\sp k}\x {L\sp m}\}=
  {\rm k m} \Tr_{E\otimes E} {L\sp {k-1}}\otimes{L\sp {m-1}} \{L\x L\}=
  0,
$$
which vanishes due to the cyclicity of the trace. By a result of 
Babelon and Viallet \cite{BV}, such
an R-matrix is guaranteed to exist if the eigenvalues of $L$ are in
involution. The Liouville integrability of a system represented by a Lax pair
has been reduced then to finding any solution to (\ref{rmatrix}) and 
counting the number of independent traces.
Further, the R-matrix is an essential ingredient when examining the
separation of variables of such integrable systems \cite{Sk, KNS}.

Unfortunately the construction of R-matrices has hitherto been somewhat of
an arcane art and many have been obtained in a case by case manner
\cite{AT}. The purpose
of this note is to present four results that address the existence and
construction of 
solutions of (\ref{rmatrix}) and hence the Liouville integrability of the
system under consideration. They
yield a uniform construction of R-matrices. In fact the R-matrices
satisfying (\ref{rmatrix}) are by no means unique and our construction
characterises this ambiguity. The approach applies equally to R-matrices with
spectral parameter.
We will illustrate these results with a simple example.
At the outset we remark that the R-matrix solutions to (\ref{rmatrix})
are generically momentum dependent. Within this family
of solutions some may be particularly simple: they may for example be
constant  (as in the Toda system \cite{OF}) or momentum independent.
We are content here with providing the construction of an R-matrix given a
Lax matrix $L$ and so answering the question of Liouville integrability:
we do not seek to further specify the momentum
or position dependence of the solution. For the 
elliptic Calogero-Moser model there are in fact \cite{BS}
no R-matrices that are independent of both momentum  and spectral parameter
(for more than 4 particles) and this illustrates the fact that simple
assumptions on the parameter dependence of an R-matrix need not be natural.
Elsewhere
we will apply these results to the elliptic Calogero-Moser models without
spectral parameter.

Our approach is as follows. First we rewrite (\ref{rmatrix}) in the form
of the matrix equation
\beq
A\sp{T}X-X\sp{T}A=B.
\label{matrixeqn}
\eeq
Here $A$ is built out of $L$ and the Lie algebra, the unknown matrix $X$ 
being solved for is essentially the
R-matrix in a given basis and $B$ represents
the right-hand side of (\ref{rmatrix}).
Our first result is to give necessary and
sufficient conditions for (\ref{matrixeqn}) to admit solutions
together with its general solution. This general solution encodes the possible
ambiguities of the R-matrix.
Because $A$ is (in general) singular  our solution is in terms of a 
generalized inverse $G$ satisfying
\beq
AGA=A\quad\quad{\rm and}\quad\quad GAG=G.
\label{geninverse1}
\eeq
Such a generalized inverse always exists. (Accounts of generalized
inverses may be found in \cite{AG,Ca,Pr,RM}.) Indeed the  Moore-Penrose inverse
-which is unique and always exists- further
satisfies $(AG)\sp\dagger=AG$, $(GA)\sp\dagger=GA$.
Observe that given  a $G$ satisfying (\ref{geninverse1}) we  have at hand
projection operators $P_1=GA$ and $P_2=AG$ which satisfy
\beq
AP_1=P_2A=A,\quad\quad P_1G=GP_2=G.
\label{proj}
\eeq
Our second result shows that the choice of generalized inverse $G$ only
alters the R-matrix within the ambiguities specified by the general solution,
and so any generalized inverse suffices to solve (\ref{matrixeqn}) and
hence construct an R-matrix.
At this stage we have reduced the problem of constructing an R-matrix to
that of constructing a generalized inverse $G$ and our third result
constructs such for a generic element $L$ of ${\mathfrak{g}}$. Because the
Moore-Penrose inverse is unique, our fourth result is to present this inverse
for generic $L$ though we shall not need to use this in our application.

The letter is organised as follows. In the next section we present the
four results. The proofs of the first two are somewhat lengthy and
algebraic and will be presented elsewhere \cite{Br1}; 
the proofs of the remaining two
are easier to outline. In section 3 we apply these to give the R-matrix
for generic $L$. 
In section 4 we extend the results to include a spectral parameter.
Section 5 is an illustrative example.
We conclude with a brief discussion.

\section{Four Results}
Our first task is to identify (\ref{rmatrix}) with (\ref{matrixeqn}).
Let  $T_\mu$ denote a  basis for the  (finite dimensional)
Lie algebra ${\mathfrak{g}}$ 
with $[T_\mu,T_\nu]=c_{\mu \nu}\sp\lambda\ T_\lambda$ defining the 
structure constants of ${\mathfrak{g}}$. Set $\phi(T_\mu)=X_\mu$,  where
$\phi$ yields the
representation $E$ of the Lie algebra ${\mathfrak{g}}$; we may take this
to be a faithful representation. With 
$L= \sum_{\mu}L\sp\mu X_\mu $ the left-hand side of (\ref{rmatrix})
becomes
$$
\{L\x L\}=\sum_{\mu,\nu} \{ L\sp\mu, L\sp\nu \} X_\mu\otimes X_\nu
$$
while upon setting $R=R\sp{\mu\nu}X_\mu\otimes X_\nu$ and
${R\sp{T}}=R\sp{\nu\mu}X_\mu\otimes X_\nu$ the right-hand side
yields
\begin{eqnarray}
[R, L\otimes 1]-[{R\sp{T}},1\otimes L]&=&
R\sp{\mu\nu}([X_\mu,L]\otimes X_\nu-X_\nu\otimes[X_\mu,L])\cr
&=& R\sp{\mu\nu}L\sp\lambda
 ([X_\mu,X_\lambda]\otimes X_\nu-X_\nu\otimes[X_\mu,X_\lambda])\cr
&=&( R\sp{\tau\nu}c_{\tau\lambda}\sp\mu L\sp\lambda -
    R\sp{\tau\mu}c_{\tau\lambda}\sp\nu L\sp\lambda) X_\mu\otimes X_\nu.
\nonumber
\end{eqnarray}
By identifying 
$ A\sp{\mu\nu}= c_{\mu\lambda}\sp\nu L\sp\lambda \equiv-ad(L)\sp\nu _\mu$,
$ B\sp{\mu\nu}=\{ L\sp\mu, L\sp\nu \}$ and $ X\sp{\mu\nu}=R\sp{\mu\nu}$
we see that (\ref{rmatrix}) is an example of (\ref{matrixeqn}).

Having shown how to identify (\ref{rmatrix}) with the matrix equation
(\ref{matrixeqn}) we may now state our first result.
\begin{thm}
The matrix equation (\ref{matrixeqn}) has solutions if and only if
$$
\displaylines{
(C1)\quad\quad\phantom{xxxx} B\sp{T}=-B,\hfill \cr
(C2)\quad\quad (1-P_1\sp{T})B(1-P_1)=0,\hfill\cr}
$$
in which case the general solution is
\beq
X={1\over2} G\sp{T} B P_1+G\sp{T} B(1-P_1) +(1-P_2\sp{T})Y+ (P_2\sp{T}ZP_2)A
\label{gensoln}
\eeq
where $Y$ is arbitrary and $Z$ is only constrained by the requirement that
$P_2\sp{T}ZP_2$ be symmetric.
\end{thm}
Although the general solution appears to depend on the  generalized
inverse $G$ we in fact find
\begin{thm}
If $\bar G$ is any other solution of (\ref{geninverse1})
with attendant projection operators $\bar P_{1,2}$ then (\ref{gensoln})
may also be written
$$
X={1\over2} {\bar G}\sp{T} B{\bar P_1}+{\bar G}\sp{T}B(1-{\bar P_1})+
  (1-{\bar P_2}\sp{T}){\bar Y}+{\bar P_2}\sp{T}{\bar Z}{\bar P_2} A
$$
where
$$
{\bar Y}= (1-P_2\sp{T})Y +P_2\sp{T}ZP_2 A+G\sp{T} B(1-{1\over2}P_1)
\quad\quad
{\bar Z}=Z+ {1\over2}(G\sp{T} B{\bar G}-{\bar G}\sp{T} B G).
$$
\end{thm}
Thus ${\bar Z}$ is again symmetric and we have a solution of the
form (\ref{gensoln}).

In the R-matrix context the matrix $B$ is manifestly antisymmetric because
of the antisymmetry of the Poisson bracket and so $(C1)$ is clearly
satisfied. We have thus reduced the existence of an R-matrix to the 
single consistency equation $(C2)$ and the construction of a generalized
inverse to $ad(L)$. We turn now to the construction of the
generalized inverse.

Let $X_\mu$ denote a Cartan-Weyl basis for the Lie algebra ${\mathfrak{g}}$.
That is $\{X_\mu\}=\{H_i,E_\alpha\}$, where $\{H_i\}$ is a basis for the 
Cartan subalgebra ${\mathfrak{h}}$
and $\{E_\alpha\}$ is the set of  step operators (labelled by
the root system $\Phi$ of ${\mathfrak{g}}$). The structure constants
are found from 
$$
[H_i,E_\alpha ]=\alpha_i E_\alpha,\quad
[E_\alpha,E_{-\alpha}]= \alpha\sp\vee\cdot H\quad 
{\rm and}\quad
[E_\alpha,E_\beta ]=N_{\alpha,\beta}E_{\alpha+\beta}
\quad{\rm if}\ \alpha+\beta\in\Phi .
$$
Here $N_{\alpha,\beta}=c_{\alpha\, \beta }^{\alpha+ \beta }$.
With these definitions we then have that
\beq
\begin{array}{rrl}
& j\qquad\quad\beta\qquad\quad&\\
&\downarrow\qquad\quad\downarrow\qquad\quad&\\
\\
\left( adL\right) =& 
\begin{array}{c}
i\rightarrow \\ 
\alpha \rightarrow
\end{array}
\left( 
\begin{array}{cc}
0 & -\beta _{i}^{\vee }L^{-\beta } \\ 
-\alpha _{j}L^{\alpha } & \Lambda _{\beta }^{\alpha }
\end{array}
\right) &=\left( 
\begin{array}{cc}
0 & u^{T} \\ 
v & \Lambda
\end{array}
\right) 
\end{array}
\eeq
where we index the rows and columns first by the Cartan subalgebra basis
$\{i,j:1\ldots \rank\mathfrak{g}\}$ then the root system 
$\{\alpha,\beta \in \Phi\}$. We will use this block decomposition of
matrices throughout. Here $u$ and $v$ are 
$|\Phi|\times \rank\mathfrak{g}$ matrices and we have introduced the
$|\Phi|\times|\Phi|$ matrix
\beq
\Lambda _{\beta }^{\alpha }=
\alpha \cdot L\, \delta_{\beta }^{\alpha }+
c_{\alpha -\beta \beta }^{\alpha }\, L^{\alpha -\beta },
\label{Lambdadef}
\eeq
where $\alpha \cdot L=\sum_{i=1}\sp{\rank\mathfrak{g}}\alpha_i L\sp{i}$.
With these definitions we have

\begin{thm} For generic $L$ the matrix $\Lambda $ is invertible and a
 generalised inverse of $adL$ is given by
\beq
\left( \begin{array}{cc}
1 & 0 \\ 
-\Lambda ^{-1}v & 1
\end{array}
\right) \left( 
\begin{array}{cc}
0 & 0 \\ 
0 & \Lambda ^{-1}
\end{array}
\right) \left( 
\begin{array}{cc}
1 & -u^{T}\Lambda ^{-1} \\ 
0 & 1
\end{array}
\right)
=\left( \begin{array}{cc}
0&0\\
0 & \Lambda ^{-1}
\end{array}
\right). 
\label{geninvadL}
\eeq
\end{thm}

We  establish the result  by first showing that for generic $L$
\beq
\left( adL\right) =
\left( \begin{array}{cc}
1 & u^{T}\Lambda ^{-1} \\
0 & 1
\end{array}
\right) \left(
\begin{array}{cc}
0 & 0 \\
0 & \Lambda
\end{array}
\right) \left(
\begin{array}{cc}
1 & 0 \\
\Lambda ^{-1}v & 1
\end{array}
\right);
\label{factadL}
\eeq
it then follows that (\ref{geninvadL}) is a
generalised inverse for $ad L$ by direct multiplication.

Now for any matrices $m$ and $\Lambda$  we have the general factorisation
\cite{Greville,AG}
$$
\left( \begin{array}{cc}
m & u^{T} \\
v & \Lambda
\end{array}
\right)  =
\left( \begin{array}{cc}
1 & u^{T}\, \Xi \\
0 & 1
\end{array}
\right) \left(
\begin{array}{cc}
m-u^{T}\, \Xi\, v & u^{T}(1-\Xi\Lambda) \\
(1-\Lambda\Xi)v & \Lambda
\end{array}
\right) \left(
\begin{array}{cc}
1 & 0 \\
\Xi\, v & 1
\end{array}
\right)
$$
where $\Xi$ is a generalised inverse of $\Lambda$.
In particular, when $m=0$ and $\Lambda$ is invertible (and so
$\Xi=\Lambda\sp{-1}$) this shows that
\beq
\left(
\begin{array}{cc}
0 & u^{T} \\
v & \Lambda
\end{array}
\right)
 =
\left( \begin{array}{cc}
1 & u^{T}\Lambda ^{-1} \\
0 & 1
\end{array}
\right) \left(
\begin{array}{cc}
-u^{T}\Lambda ^{-1}v & 0 \\
0 & \Lambda
\end{array}
\right) \left(
\begin{array}{cc}
1 & 0 \\
\Lambda ^{-1}v & 1
\end{array}
\right).
\label{interfac}
\eeq
Thus (\ref{factadL}) and hence the result follow by establishing
that $\Lambda$ is generically invertible and that
\beq
 u^{T}\Lambda ^{-1}v =0 .
\label{ulv}
\eeq

From (\ref{Lambdadef}) we see that $\Lambda $ is the perturbation
of a diagonal matrix and so is generically invertible:
the zero locus $\det\Lambda=0$ is a polynomial in the coefficients
of $ad L$ and so the complement of this set is dense and open.
For such an invertible $\Lambda $ we thus have
\beq
\rank \Lambda= \dim\Lambda=\dim\mathfrak{g}- \rank\mathfrak{g}.
\label{rank}
\eeq
Now the maximum rank\footnote{
If $\det(t-ad L)=\sum_{j=0}\sp{\dim\mathfrak{g}} p_j(L)\,t\sp{j}$
is the characteristic polynomial of $ad L$, the regular semi-simple elements
of a semi-simple Lie algebra $\mathfrak{g}$ are those elements for which 
$p_{\rank\mathfrak{g}}(L)\ne 0$. These elements
are also of rank $\dim\mathfrak{g}- \rank\mathfrak{g}$ and form an open dense
set in $\mathfrak{g}$, but this condition is different from
$\det\Lambda\ne0$.} of the
matrix $ad L$ is $\dim\mathfrak{g}- \rank\mathfrak{g}$ \cite{Varadarajan}.
From (\ref{interfac})  we see that
$\rank \Lambda+\rank(u^{T}\Lambda ^{-1}v)
= \rank (ad L)\leq\dim\mathfrak{g}-\rank\mathfrak{g}$
 and so from (\ref{rank}) we deduce that $\rank \Lambda=\rank (ad L)$.
Therefore $\rank(u^{T}\Lambda ^{-1}v)=0$ and consequently
(\ref{ulv}) must hold. The result then follows.

An alternate factorisation of $ad L$ is possible
for the generic $L$ under consideration.
Utilising (\ref{ulv}) we find that
$$
(ad L)= 
\left(\begin{array}{c} u\sp{T} \Lambda ^{-1}\\ 1 \end{array}\right)
\Lambda
\left(\begin{array}{cc} \Lambda ^{-1}v& 1 \end{array}\right)
=E \,\Lambda\, F.
$$
Employing a result of MacDuffee \cite{AG} this full rank 
factorisation then yields:

\begin{thm} For generic $L$ the Moore-Penrose inverse of $ad L$
is given by
$$
F\sp{\dagger} (F\,F\sp{\dagger})\sp{-1} \Lambda ^{-1}
(E\sp{\dagger}\, E)\sp{-1} E\sp{\dagger}
$$
where $E=\left(\begin{array}{c} u\sp{T} \Lambda ^{-1}\\ 1 \end{array}\right)$
and   $F=\left(\begin{array}{cc} \Lambda ^{-1}v& 1 \end{array}\right)$.
\end{thm}

\section{The R-matrix}
We now bring together the results of the previous section to present the
R-matrix for a generic $L$ when this exists.
From the fact that $A=-(ad L)\sp{T}$ a generalized inverse of $A$
is given by minus the transpose of the  generalized inverse (\ref{geninvadL}).
Utilising our earlier notation this means that we have the  projectors
$$
P_{1}=\left(
\begin{array}{cc}
0 & 0 \\
\Lambda ^{-1T}u & 1
\end{array}
\right)
,\quad\quad
P_{2}=\left(
\begin{array}{cc}
0 & v ^{T}\Lambda ^{-1T} \\
0 & 1
\end{array}
\right).
$$
Let us express the Poisson brackets of the entries of $L$ in the same
block form in the Cartan-Weyl basis:
$$
B=\left(
\begin{array}{cc}
\zeta & - \mu\sp{T} \\
\mu & \phi 
\end{array}
\right)= -B ^{T}
$$
where $B\sp{\alpha j }=\{L\sp\alpha,L\sp{j}\}=\mu_{\alpha j }$ and so on.
The constraint $(C1)$ is manifestly satisfied.

The constraint $(C2)$ is now (the $\rank\mathfrak{g}\times\rank\mathfrak{g}$
matrix equation)
\beq
(C2)\quad\quad\quad\quad\quad\quad
0=\zeta +\mu\sp{T} \Lambda ^{-1T} u - u\sp{T}\Lambda ^{-1} \mu +
u\sp{T}\Lambda ^{-1}\phi\, \Lambda ^{-1T} u.
\quad\quad\quad\quad\quad
\eeq
Each term in this equation is known and so the equality may be readily
checked.

Supposing the constraint $(C2)$ is satisfied we then find
from (\ref{gensoln}) the general R-matrix takes the form
\beq
R=\left(
\begin{array}{cc}
0 & 0 \\
- \Lambda ^{-1}\mu +\frac{1}{2} \Lambda ^{-1}\phi\, \Lambda ^{-1T} u
& -\frac{1}{2} \Lambda ^{-1}\phi
\end{array}
\right) +\left(
\begin{array}{cc}
p & q \\
-\Lambda ^{-1}v p-Fu & -\Lambda ^{-1}v q-F\Lambda ^{T}
\end{array}
\right).
\label{rmat}
\eeq
The second term characterises the ambiguity in R where we have
parameterised the matrices $Y,Z$ in (\ref{gensoln}) by
$Y=\left(\begin{array}{cc}p&q\\ r&s \end{array} \right)$
and $Z=\left(\begin{array}{cc}a&b\\ c&d \end{array} \right)$.
Here the matrices $p,q$ are arbitrary 
while the entries of $Z$ are such that
\beq
F=\Lambda ^{-1}v a v ^{T}\Lambda ^{-1T}+d+\Lambda
^{-1}v b+c v ^{T}\Lambda ^{-1T}
\label{Fdef}
\eeq
is symmetric.

\section{Inclusion of Spectral Parameter}
For simplicity we have presented our construction for Lax pairs with
no spectral parameter but it is straightforward to incorporate such
a parameter.
The relevant equation to be solved for is now
\beq
\{L(u)\x L(v)\}=[R(u,v), L(u)\otimes 1]-[{R\sp\pi}(u,v),1\otimes L(v)],
\label{rmatrixsp}
\eeq
where if $R(u,v)=R\sp{\mu\nu}(u,v)X_\mu\otimes X_\nu$ then
${R\sp\pi}(u,v)$  is defined by\footnote{We use $\pi$ to denote
both matrix transposition together with the interchange of $u$ and $v$
while $T$ denotes ordinary matrix transposition.}
${R\sp\pi}(u,v)=R\sp{\nu\mu}(v,u)X_\mu\otimes X_\nu$.
Now
$$
B\sp{\mu\nu}(u,v)=\{L\sp{\mu}(u),L\sp{\nu}(v)\}=-B\sp{\nu\mu}(v,u)
$$
and because $L(u)$ depends on $u$ alone the generalized inverse
now also depends on the spectral parameter as $G=G(u)$.
The  equation we now wish to solve is
$$
A\sp{T}(u)X(u,v)-X\sp{\pi}(u,v)A(v)=
A\sp{T}(u)X(u,v)-X\sp{T}(v,u)A(v)=B(u,v),
$$
and this has the analogous solution
$$
\begin{array}{rl}
X(u,v)=&{1\over2} G\sp{T}(u) B(u,v) P_1(v)+G\sp{T}(u) B(u,v)(1-P_1(v))\\
& +( 1-P_2\sp{T}(u))Y(u,v)+ (P_2\sp{T}(u)Z(u,v)P_2(v))A(v)
\end{array}
$$
if and only if
$$
\displaylines{
(C1\sp\prime)\quad\quad\phantom{xxxx} B\sp{\pi}=-B,\hfill \cr
(C2\sp\prime)\quad\quad (1-P_1\sp{T}(u))B(u,v)(1-P_1(v))=0.\hfill\cr}
$$
Here $Y(u,v)$ is arbitrary while the symmetry condition now becomes
$$(P_2\sp{T}(u)Z(u,v)P_2(v))\sp\pi = P_2\sp{T}(u)Z(u,v)P_2(v).$$
As in the spectral parameter independent case, this reduces to the
requirement that
\beq
\begin{array}{rl}
F(u,v)=&\Lambda ^{-1}(u)v(u) a(u,v) v ^{T}(v)\Lambda ^{-1T}(v)+d(u,v)+
\Lambda ^{-1}(u)v(u) b(u,v)\\
&+c(u,v) v ^{T}(v)\Lambda ^{-1T}(v)
\end{array}
\label{Fdefsp}
\eeq
be such that $F\sp\pi=F$.

\section{An Example}
We conclude with the simple but illustrative example of the 
harmonic oscillator presented as the Lax pair (with spectral parameter)
\beq
L(u)=\left(
\begin{array}{cc}
ipx/u & (p\sp2 /u) +1 \\
(x\sp2/u) +1& -ipx/u
\end{array}
\right), 
\quad\quad
M(u)=\left(
\begin{array}{cc}
0&i\\
i&0
\end{array}
\right).
\eeq
The consistency of the Lax equation 
$\dot L(u) =\left[ L(u),M(u)\right] $ follows from the equations of motion
of the Hamiltonian $H=(p\sp2 +x\sp2 +u)/2 = -(u/2) \det L(u)$.

Although we could equally work with the simple algebra $su(2)$ in this example
we will take the algebra to be $gl(2)$. Now for any $l\in gl(2)$,
$$
l=\left(
\begin{array}{cc}
a&b\\
c&d\end{array}
\right)=
a H_1 +b E_{12} +c E_{21}+ d H_2,
$$
we find that in our Cartan-Weyl basis
$$
ad\ l=\left(
\begin{array}{rrrr}
0&0&-c&b\\
0&0&c&-b\\
-b&b&a-d&0\\
c&-c&0&d-a
\end{array}
\right).
$$
Then
$$
u=\left(
\begin{array}{rr}
-c&c\\
b&-b
\end{array}
\right),
\quad
v=\left(
\begin{array}{rr}
-b&b\\
c&-c
\end{array}
\right),
\quad
\Lambda=\left(
\begin{array}{cc}
a-d&0\\
0&d-a
\end{array}
\right)
$$
and $l$ is generic\footnote{It is regular semi-simple provided
$(a-d)\sp2 +bc\ne0$.} provided $a-d\ne0$.
We note that in the Cartan-Weyl basis the permutation operator 
$P=\sum_{i} H_i\otimes H_i+\sum_{\alpha\in \Phi} E_{\alpha}\otimes E_{-\alpha}$
(which is such that $P( X\otimes Y) P=Y\otimes X$) takes the form
\beq
P=\left(
\begin{array}{cccc}
1&0&0&0\\
0&1&0&0\\
0&0&0&1\\
0&0&1&0
\end{array}
\right).
\label{permP}
\eeq

Now for the case at hand $b=(p\sp2/u)+1$, $c=(x\sp2/u)+1$,
$a=-d=i p x/u$ and $L$ is generic for $px\ne0$. We may therefore use the
expressions computed in sections 3 and 4. We calculate (using $\{p,x\}=1$)
that
$$
B(u,v)=\frac{1}{uv}\left(
\begin{array}{rrrr}
0&0&-2ip\sp2&2ix\sp2\\
0&0&2ip\sp2&-2ix\sp2\\
2ip\sp2&-2ip\sp2&0&4xp\\
-2ix\sp2&2ix\sp2&-4xp&0
\end{array}
\right)
=
\left(
\begin{array}{rr}
0&-\mu\sp{T}(u,v)\\
\mu(u,v)&\phi(u,v)
\end{array}
\right).
$$
and straightforwardly verify that condition $(C2\sp\prime)$ is satisfied.
The R-matrix is then given by
$$
R(u,v)=\left(
\begin{array}{cc}
0 & 0 \\
- \Lambda ^{-1}(u)\mu(u,v) +\frac{1}{2} \Lambda ^{-1}(u)\phi(u,v)\,
 \Lambda ^{-1T}(u) u(v)
& -\frac{1}{2} \Lambda ^{-1}(u)\phi(u,v)
\end{array}
\right)
$$
$$
 +\left(
\begin{array}{cc}
p(u,v) & q(u,v) \\
-\Lambda ^{-1}(u)v(u) p(u,v)-F(u,v)u(v) & -\Lambda ^{-1}(u)v(u) q(u,v)-F(u,v)
\Lambda ^{T}(v)
\end{array}
\right).
$$
The second term again characterises the ambiguity in R and we have
parameterised the matrices $Y(u,v)$ and $Z(u,v)$ in an analogous way to
the spectral parameter independent case of section 3.
Substitution of the various quantities gives for the first term
$$
R(u,v)=
\left(
\begin{array}{cccc}
0&0&0&0\\
0&0&0&0\\
\frac{-p\sp2+v}{2pxv}&-\frac{-p\sp2+v}{2pxv}&0&\frac{i}{v}\\
\frac{-x\sp2+v}{2pxv}&-\frac{-x\sp2+v}{2pxv}&\frac{i}{v}&0
\end{array}
\right).
$$
This R-matrix is clearly dynamical.
Making use of the the block structure of the R-matrix we see that by choosing
$$
p(u,v)=\frac{-2i}{u-v}\left(
\begin{array}{cc}
1&0\\
0&1
\end{array}\right),
\quad
q(u,v)=0,
\quad
F(u,v)=-\frac{u+v}{u-v}\frac{1}{2px}\left(
\begin{array}{rr}
0&1\\
-1&0
\end{array}\right)
$$
we arrive at the nondynamical
$$
R(u,v)=\frac{-2i}{u-v}\, P
$$
where $P$ is given by (\ref{permP}).

\section{Discussion}
We have presented a uniform construction for a classical R-matrix given
a Lax pair, thus answering the question of the Liouville integrability
of the system in terms of the invariants of the matrix $L$. The method
not only gives necessary and sufficient conditions for the R-matrix to exist
and describes its ambiguities, but is algorithmic as well.
Given $L$, first construct $ad L$. Next construct any generalized inverse
to $ad L$ and verify $(C2)$; this is the necessary and sufficient condition
for an R-matrix to exist: it is given explicitly  by
(\ref{gensoln}). Further we have given a generalized inverse for generic $L$
in (\ref{geninvadL}); genericity is easily checked by evaluating
$\det \Lambda\ne0$, where $\Lambda$ is the restriction of $ad L$ to the
root space (given by (\ref{Lambdadef})).
The ambiguities in the R-matrix have been specified. We remark, that
the block nature of the R-matrix allows us to easily verify putative ansatz
for a given R-matrix.

Thus far our discussion has been limited to linear R-matrices and we
briefly discuss the application to quadratic r-matrices, ie. the solutions
to
\beq
\{L\x L\}=[r,L\otimes L]=[r_A,L\otimes L],
\label{quadr}
\eeq
where $r_A=(r-r\sp{T})/2$. (It follows from the anti-symmetry of $\{,\}$ that
$[r+r\sp{T},L\otimes L]=0$.)
As discussed in \cite{BV}, the quadratic R-matrix calculation may be 
reduced to the linear R-matrix situation. In particular the R-matrix
\beq
R=\frac{1}{2}r_A\sp{\mu\nu}L\sp\lambda\left(X_\mu\otimes X_\nu X_\lambda
+X_\mu \otimes  X_\lambda X_\nu 
\right)
\label{quadreqn}
\eeq
that satisfies (\ref{rmatrix}) yields a solution $r_A$ of (\ref{quadr});
the general solution is then built from $r_A$ and the centraliser of
$L\otimes L$.
Our theorem has given us the left-hand side of
(\ref{quadreqn}) and a quadratic  r-matrix is then given by solving the linear
equation
$R\sp{\mu\sigma}=r_A\sp{\mu\nu}F_\nu\sp{\sigma}$ where	
$F_\nu\sp{\sigma}=(F_{\nu\lambda}\sp{\sigma}+
   F_{\lambda\nu}\sp{\sigma})L\sp\lambda/2$
and $X_\nu X_\lambda=F_{\nu\lambda}\sp{\sigma}\, X_\sigma$.
Whereas the linear R-matrix involves only Lie algebraic data, the
quadratic r-matrix may involve the group structure through the
multiplication $X_\nu X_\lambda=F_{\nu\lambda}\sp{\sigma}\, X_\sigma$.
Nonetheless, the quadratic r-matrix has been reduced to a linear equation
amenable to direct solution.

Finally we mention that for systems obtained by Hamiltonian reduction
an alternative geometric construction  of classical R-matrices exists 
\cite{ABT} in terms of Dirac brackets. This suggests there is a correspondence
between Dirac brackets and generalized inverses. This is indeed the case and
I will present this elsewhere.

\section{Acknowledgements}
This material was presented at the CRM \lq Workshop on Calogero-Moser-
Sutherland Models\rq\ (Montreal, March 1997) and I thank 
the organisers and participants for such a stimulating meeting.
I have benefited from comments by
J. Avan, J. Harnad, A.N.W. Hone,
I. Krichever, V. Kuznetsov, M. Olshanetsky and E. Sklyanin.

\end{document}